\documentclass{jetpl}
\twocolumn

\lat

\title{Induced Absorption Resonance on the Open
$F_g = 1 \rightarrow F_e = 2$ Transition of the $D_1$ Line of the
$^{87}{Rb}$ Atom}

\rtitle{Induced absorption resonance on the open \ldots}

\sodtitle{Induced Absorption Resonance on the Open $F_{g} = 1
\rightarrow F_{e} = 2$ Transition of the $D_1$ Line of the
$^{87}{Rb}$ Atom}

\author{$A. S. Zibrov^{a,b}$ and $A. B. Matsko^{c}$}

\rauthor{A. S. Zibrov, A. B. Matsko}

\sodauthor{Zibrov, Matsko}

\address{$^{a}$ Lebedev Physical Institute, Russian Academy of Sciences, Leninsky
pr. 53, Moscow, 117924, Russia,\\ $^{b}$ Department of Physics,
Harvard University, Cambridge, Massachusetts, 02138, USA,\\
e-mail:
{azibrov@cfa.harvard.edu} \\
$ ^{c}$ Jet Propulsion Laboratory, California Institute of
Technology, 4800 Oak Grove Drive, Pasadena, California,
91109-8099, USA}

\dates{1 August 2005}

\abstract{Induced absorption resonance on the open $F_g=1
\rightarrow F_e = 2$ transition of the $D_1$ line of the $^{87}$
Rb atom has been observed. The effect of atomic motion on the
formation of the resonance has been revealed. The numerical
calculations are in good agreement with experiment.}


\begin{document}

\maketitle

Narrow atomic resonances are always important in spectroscopy,
particularly in metrology. A new type of resonances was recently
discovered in a degenerate two-level system and was called induced
absorption resonance \cite{1}. Rautian \cite{2} pointed out that
spontaneous coherence transfer plays an important and universal
role in the formation of spectra. If transitions with
approximately equal frequencies $\nu_{m_1-n_1}$ and $\nu_{m-n}$,
where $m_i$ and $n_i$ are the upper and lower levels,
respectively, are coupled through spontaneous relaxation, it
transfers coherence between the states $| m_i\rangle$ and $|
n_i\rangle$. Rautian considers that the importance of this process
is comparable with the importance of the spontaneous and
stimulated emission processes introduced by Einstein. In the
spectroscopy of a coherently prepared medium, coherence
spontaneous transfer is clearly manifested when induced absorption
resonance is observed. Its name appeared by analogy with induced
transparency resonance. In the case of induced absorption
resonance, absorption increases due to constructive interference
between the quantum states of the system. Coherence is transferred
between Zeeman sublevels due to a spontaneous process \cite{3,4}.
As was recently predicted in \cite{5}, spontaneous coherence
transfer could be efficiently used in the $\gamma$ radiation range
to control Mossbauer spectra. Analysis of spontaneous coherence
transfer seems to be important, because this effect is the same in
microwave and optical ranges, as well as in $\gamma$ optics. This
motivation stimulated us to study induced absorption resonance in
the rubidium vapor and to present the results in this paper.

Induced absorption resonance was first detected on the closed
transition of the $D_2$ line of the Rb atom that absorbs two
in-phase copropagating light waves \cite{1}. It was pointed out
that induced absorption resonance was observed when the degeneracy
factor of the excited state exceeded the degeneracy factor of the
ground state, i.e., $0 < F_g \leq F_e = F_g + 1$ \cite{6}. This
effect was theoretically described in \cite{7} for various
intensities of the pump field, magnetic moments, and
polarizations.


\begin{figure}
\centering
\includegraphics[height=5cm]{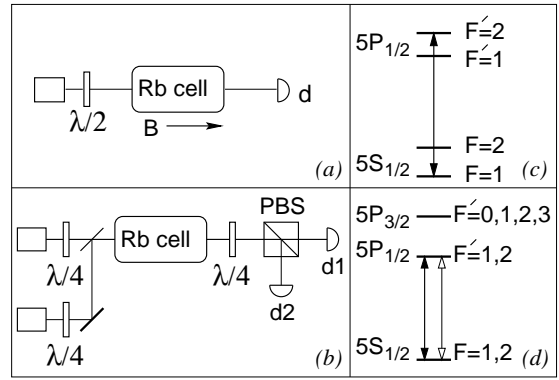}
\caption{ \label{1} Fig.1:~Layouts of the experiments and the
scheme of levels for observing induced absorption resonance. (a)
The Hanle configuration experiment, where the transmittance of the
linearly polarized coherent light is detected as a function of the
magnetic field B. (b) The experiment that involves a pump laser
inducing atomic coherence and the probe laser. The transmission of
a weak circularly polarized probe field is measured as a function
of the frequency. The frequency of the pump laser is fixed. The
rotation of the polarization of the pump field is opposite to the
rotation of the polarization of the probe field. (c) The scheme of
the levels of the $D_1$ line of the $^{87}$Rb atom. (d) The scheme
of the levels of the $D_1$ and $D_2$ lines of the $^{87}$Rb atom.
%
}
\end{figure}

Signals associated with induced absorption resonance were also
observed in experiment \cite{8}, where the Hanle configuration was
used (laser light propagated along the magnetic-field direction).
In that experiment, the atomic Rb vapor was pumped by a single
linearly polarized wave. In the case of the degeneracy of the
lower level of the ground state, bright resonances were detected
in fluorescence with a concomitant increase in absorption. Those
experimental results were theoretically analyzed in \cite{9}. In
the more recent work \cite{10}, it was shown that an increase in
absorption in the Hanle configuration should also be expected when
a laser beam is perpendicular to the magnetic field.

As was mentioned above, induced absorption resonance was observed
on the closed transition of the degenerate ground state. The weak
induced absorption resonance was also observed on the open $F_g =
2\rightarrow F_e = 2,3$  transition in the $^{85}$Rb atom
\cite{8}. However, on the other open transition $F_g =
1\rightarrow F_e = 2$ of the $D_1$ line of the $^{87}$Rb atom, the
effect was not detected \cite{11,12}. It was assumed that this was
due to optical pumping and the low degeneracy factor of the
corresponding atomic states.

\begin{figure}
\centering
\includegraphics[height=8cm]{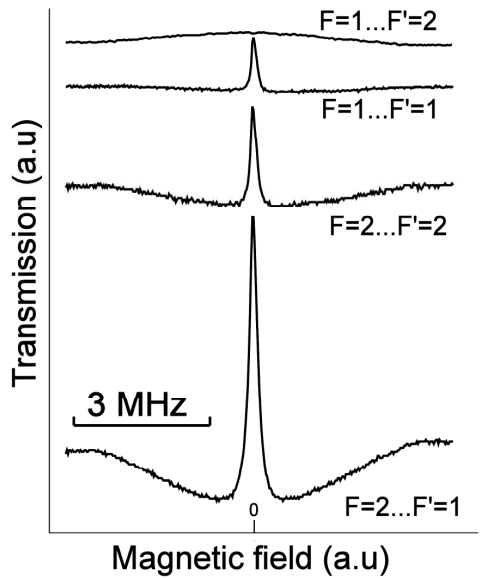}
\caption{ \label{2} Fig.2:~Transmission of laser radiation vs. the
longitudinal field magnitude (see the layout in Fig. \ref{1}a).
The laser frequency is tuned to the $D_1$ line of the $^{87}$Rb
atom. Transmission increases at zero magnetic field for all the
transitions except for the $F = 1 \rightarrow F ' = 2$ transition
(see Fig. \ref{1}c). The relative amplitudes of the resonances are
not changed. The magnitude of the induced absorption resonance on
the $ F = 1\rightarrow F' = 2$ transition is very small and is
equal to 0.2\% of the total 60\% absorption. The shape of the
resonance is shown in Fig. \ref{3}.
%
}
\end{figure}

In this paper, we present the results of the experimental
investigation of induced absorption resonance on the open $F_g = 1
\rightarrow F_e = 2$ transition of the $D_1$ line scheme \cite{1}
and in the Hanle configuration \cite{8}. The observed effect
appeared to be weak (0.2\% of the total absorption due to the
optical-pumping-induced depletion of the population). Optical
pumping does not completely destroy spontaneous coherence
transfer, which is responsible for the formation of induced
absorption resonance, because the atoms interact with light for a
finite time. Thus, this experiment corroborates that induced
absorption resonance occurs on all $F_g \rightarrow F_e = F_g + 1$
transitions both closed and open. The numerical calculation
confirms the conclusions drawn using those experimental results.

We describe both experiments. The first experiment is the same as
in \cite{11}. Figure \ref{1}a shows the layout of the setup. An
external-cavity laser was tuned to the $D_1$ or $D_2$ line of the
$^{87}$Rb atom (see Figs. \ref{1}c,\ref{1}d). A laser beam passed
through a half-wave plate and a cell 3.0 cm in length that
contained isotopically pure $^{87}$Rb. The vapor density was
controlled by the cell temperature. The transmission was detected
by a photodiode $d$. The cell was placed in a three-layer magnetic
screen. The longitudinal magnetic field is produced by a solenoid
placed inside the screen. The static magnetic field gives rise to
the appearance of Zeeman sublevels. The splitting was equal to the
splitting $\mu_B B /\hbar$  between the neighboring Zeeman
sublevels, where $\mu_B$ is the Bohr magneton. In the case of the
$^{87}$Rb atom, the splitting is equal to $B\times 0.7 MHz/G$.

\begin{figure}
\centering
\includegraphics[height=6cm]{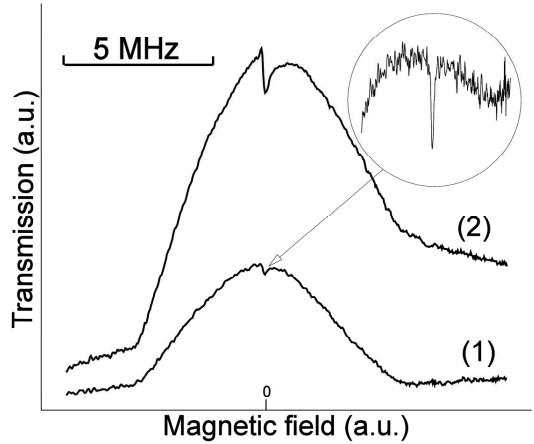}
\caption{\label{3} Fig.3:~ Transmission of laser radiation vs. the
magnetic field B in the experiment whose layout is shown in Fig.
\ref{1}a. The laser is tuned to the open $F = 1\rightarrow F' = 2$
transition (of the $D_1$ line). Transmission increases at zero
magnetic field. Lines 1 and 2 correspond to the laser-field
intensities 1.5 and 3 mW/$cm^2$, respectively. A decrease in
absorption on line 1 is equal to 3\% of the total 60\% absorption.
The inset shows the fine structures of induced absorption
resonance.
}
\end{figure}

In the second experiment, two external-cavity lasers were used.
The frequency of the laser creating coherence remained unchanged,
whereas the frequency of the probe laser was scanned. The
radiation of the strong laser after the passage through a
quarter-wave plate became circularly polarized in the direction
$\sigma^+$ , whereas the radiation of the probe laser was
polarized in the opposite direction $\sigma^-$. Downstream of the
cell, beams were split by a quarter-wave plate and polarization
cube (PBS) and were detected by photodetectors $d1$ and $d2$.

The magnetic-field dependence of the light transmission at four
frequencies in the Hanle experiment is shown in Fig.\ref{2}. The
cell temperature was equal to $50^°$ C, the light power was 0.1
mW, and the beam diameter was equal to 1.5 mm. The transmission
increases near the region where the magnetic field is zero for all
transitions except for the $F = 1\rightarrow F' = 2$ transition.
The absorption resonance with the subnatural width of the optical
transition is almost unseen in the curve. This behavior was
observed in \cite{11}.

\begin{figure}
\centering
\includegraphics[height=6cm]{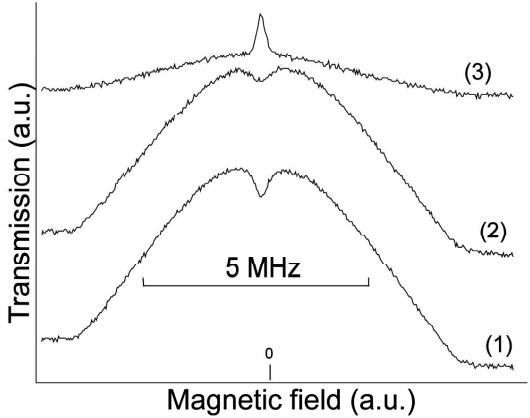}
\caption{\label{4} Fig.4:~ Transmission of laser radiation vs. the
magnetic field B in the experiment whose layout is shown in Fig.
\ref{1}a (Hanle configuration). The laser is tuned near the
transition $5S_{1/2}, F = 1\rightarrow 5P_{3/2}, F' = 0, 1, 2$ of
the $D_2$ line of the rubidium atom: (1) the laser frequency is
shifted by 250 MHz from the center of the Doppler profile toward
the blue end, (2) the laser frequency is turned to the center of
the Doppler profile, and (3) the laser frequency is shifted by 250
MHz from the center of the Doppler profile toward the red end. A
decrease of absorption at the center of line 1 is equal to 2\%,
whereas the total absorption is equal to about 60\%.
}
\end{figure}

After an increase in sensitivity, we detected this increase in
absorption (see Fig. \ref{3}). The induced absorption resonance is
broadened as the intensity of light increases and disappears at
intensities above $20 mW/cm^2$ . The total increase in absorption
is a consequence of the depopulation of levels due to optical
pumping.

An increase in absorption was also observed on the transition $5
S_ {1/2} ; F = 1 \rightarrow 5P_{3/2}; F' = 0, 1, 2 $ of the $D_2$
line of rubidium. Unfortunately, the Doppler broadening prevents
the resolution of all the transitions interacting with light in
this transition. We measured the transmission for three different
tunings of the laser frequency (see Fig. \ref{4}). An increase in
absorption is easily seen with tuning to the high-frequency part
of the Doppler profile. We emphasize that interaction with the $F
= 1 \rightarrow F' = 2 $ transition is stronger at these
frequencies, whereas interaction with the $F = 1\rightarrow F' =
0$ transition is stronger in the low-frequency range.

Transmission of laser radiation vs. the magnetic field B in the
experiment whose layout is shown in Fig. 1a (Hanle configuration).
The laser is tuned near the transition $5S_{1/2}, F = 1
\rightarrow 5P_{3/2}, F' = 0, 1, 2$ of the $D_2$ line of the
rubidium atom: (1) the laser frequency is shifted by 250 MHz from
the center of the Doppler profile toward the blue end, (2) the
laser frequency is tuned to the center of the Doppler profile, and
(3) the laser frequency is shifted by 250 MHz from the center of
the Doppler profile toward the red end. A decrease of absorption
at the center of line 1 is equal to 2\%, whereas the total
absorption is equal to about 60\%.

\begin{figure}
\centering
\includegraphics[height=6cm]{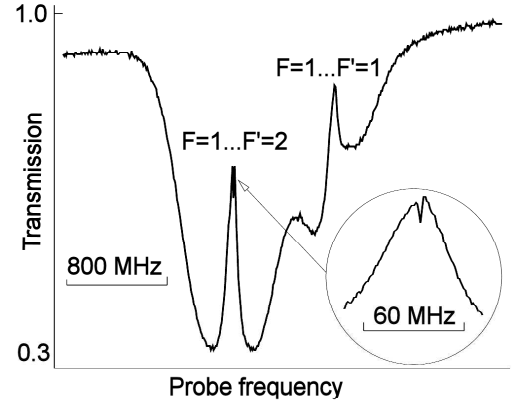}
\caption{\label{5}Fig.5:~Transmission $I_{out}/I_{in}$ of the
probe beam in the experiment whose layout is shown in Fig.
\ref{1}b. The frequency of the laser inducing coherence is tuned
to the open transition $5S_{1/2}, F = 1\rightarrow 5P_{1/2}, F' =
2$ (see Fig. \ref{1}d). The inset shows the resonance peak at an
increased scale.
}
\end{figure}

We studied induced absorption resonance in the "standard scheme"
of experiments with induced absorption, where two lasers were
used. The frequency dependence of the transmission of the
radiation of the probe laser is shown in Fig. \ref{5}. The
frequency of the laser creating coherence was tuned to the center
of the Doppler profile of the $F_g = 1 \rightarrow F' = 2$
transition. The power of this laser was equal to 3 mW, whereas the
power of the probe laser was equal to 0.1 mW. The beams of both
lasers had a diameter of 1.5 mm. Induced absorption resonance was
located at the peak of the bell-shaped transparency resonance
associated with optical pumping.

\begin{figure}
\centering
\includegraphics[height=4.5cm]{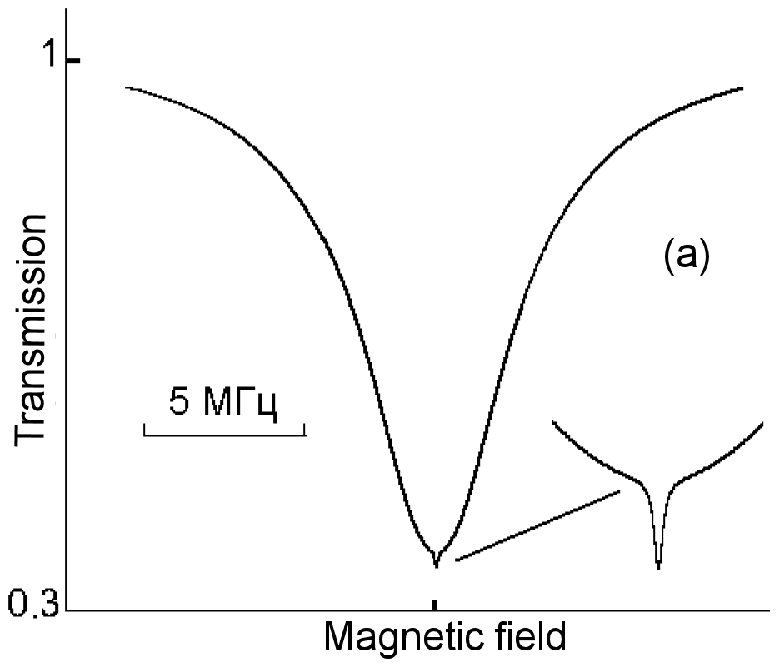} \includegraphics[height=4.5cm]{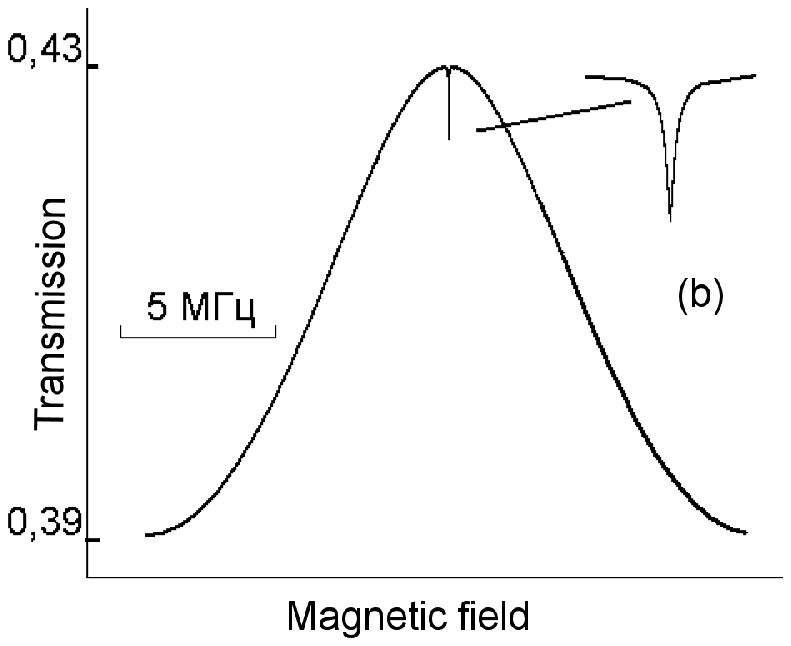}
\caption{\label{6} Fig.6:~Transmission $I_{out}/I_{in}$ on the
open $F = 1\rightarrow F' = 2$ transition vs. the magnetic field
as obtained in the numerical calculation for (a) homogeneously
broadened and (b) Doppler broadened rubidium vapor. The
calculation was performed for the case shown in Fig. \ref{1}a.
}
\end{figure}

In order to explain the experimental results, numerical
calculations based on the Maxwell-Bloch equations describing the
propagation of the electromagnetic field in the cell are performed
with the inclusion of all the Zeeman sublevels of the $D_1$ line
of the $^{87}$Rb atom. We studied the interaction with linearly
polarized light tuned to the frequency of the $F_g = 1 \rightarrow
 F_e = 2$ transition of the $D_1$ line. Light in the model
propagates in the direction coinciding with the direction of the
external magnetic field. The model also takes into account that
atoms are continuously refreshed in the region of interaction with
light due to influx from other regions of the cell. The rate
$\gamma_0$ of this process is determined by the time of flight of
atoms through the laser beam. The Bloch equations are the same as
in \cite{9}. The calculations were performed with $\gamma_0 =
0.0004\gamma$, where $\gamma$ is the rate of the natural decay of
the excited state, and the Rabi frequency $\Omega = 0.04\gamma$,
which corresponds to the field intensity $I_{in} = 0.025
mW/{cm^2}$. The normalized magnetic field dependence of the
transmission of the incident radiation is shown in Fig. \ref{6}.
Induced absorption resonance is seen near zero magnetic field. We
emphasize that induced absorption resonance on this transition was
not observed in the theoretical work \cite{9}.

As the intensity of the laser field increases, the contrast of
induced absorption resonance decreases. Moreover, we observe in
experiments that induced absorption resonance completely
disappears at intensities of several milliwatts per centimeter
squared. To understand this observation, we performed numerical
simulation along the Doppler profile with $\gamma_0 = 0.003\gamma$
and $\Omega = 0.34\gamma$, which corresponds to an intensity of
$I_{in} = 1.5 mW/cm^2$. The width of the velocity distribution was
taken to be $100\gamma$. The result of the numerical simulation is
shown by the right line in Fig. 6, where it is seen that the
behavior of the calculated line coincides with the experimentally
observed dependence given by line 1 in Fig. \ref{3}.

\begin{figure}
\centering
\includegraphics[height=6cm]{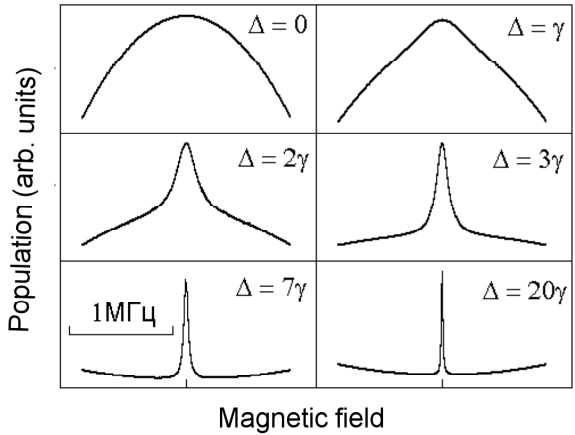}
\caption{\label{7} Fig.7:~ Magnetic-field dependence of the
population of the excited state for various detunings $\Delta$ for
the homogeneously broadened $F = 1 \rightarrow F' = 2$ transition.
As the detuning $\Delta$ increases, the population decreases. The
maximum population at $B = \Delta = 0$ is equal to 0.002. The
maximum population for the detuning $\Delta = \gamma,~2\gamma,~
3\gamma, ~7\gamma$, and $~20\gamma$ is equal to 0.98, 0.95, 0.89,
0.59, and 0.15, respectively. The intensity of the laser field and
field–atom interaction are the same as in Fig. \ref{6}b. No
subnatural width of resonance is observed for atoms with zero
velocity.
}
\end{figure}

The resulting differences shown in Figs. \ref{6}a and \ref{6}b are
obviously caused by the motion of atoms. In order to ensure this
conclusion, we calculated the population of the $F_e = 2$ excited
state for the inhomogeneously broadened transition with various
detunings $\Delta$ of the laser frequency (in other words, for
groups of atoms with various velocities). The calculations were
performed with the same Rabi frequencies and coherence decay rates
as in the calculations presented in Fig. \ref{6}b and in the
experiment (see Fig. \ref{3}). As is seen in Fig. \ref{7}, induced
absorption resonance on the homogeneously broadened transition is
not observed at zero detuning and only atoms with nonzero velocity
contribute to the formation of the resonance. Such a behavior is
due to the fact that atoms with lower velocities more rapidly
leave the process of the formation of induced absorption resonance
on the inhomogeneously broadened transition. Under the action of
optical pumping, atoms efficiently decay into another level of the
ground state and thereby the effect of spontaneous coherence
transfer, which determines the formation of induced absorption
resonance, is small.

Spontaneous coherence transfer is substantial for the $F_g
\rightarrow F_e = F_g + 1$ transition, because the population of
the magnetic sublevels of the excited state exceeds the population
of the sublevels of the states of the $F_g \rightarrow F_e = F_g$
or $F_g\rightarrow F_e = F_g -1$ transitions under the same
conditions. Atoms in these states are trapped in dark states due
to coherent population trapping \cite{13}, whereas this is not the
case for the $F_g\rightarrow F_e = F_g + 1$ transition.

It is also easy to explain why absorption on the induced
absorption resonance near the open transition is small in our
experiment. Near the open transition, atoms efficiently decay into
another level of the ground state and thereby the effect of
spontaneous coherence transfer, which determines the formation of
induced absorption resonance, is small.

Thus, induced absorption resonance on the open $F_g =1 \rightarrow
F'_e = 2$ transition of the $D_1$ line of rubidium has been
observed in the experiment. Experiments show that induced
absorption resonance is observed even in the presence of strong
optical pumping on the transition. Therefore, the previous
statement made in \cite{1} that a closed transition is necessary
for observing induced absorption resonance should be revised. This
study also shows that the intensity range for observing induced
absorption resonance in a Doppler broadened medium is wider than
that in a homogeneously broadened medium.

In conclusion, we try to explain why the induced absorption
resonance on the open transition under investigation was not
observed in the previous theoretical and experimental works cited
above. The population of the open transition is equal to zero in
the steady state. In a real experiment, fresh atoms from other
regions of the cell enter the region of atom-light interaction.
This process maintains a nonzero population of the state. The
shorter the interaction time, the larger the number of atoms in
the excited state. However, when the interaction time is too
short, coherence degrades. It is necessary to search for the
optimum interaction time. The interaction time is determined by
the laser-beam radius. An increase in the radius increases the
interaction time. In our experiment, we chose the beam radius
according to the available intensities. We think that optimization
determined the success in the observation of the induced
absorption resonance in this experiment.

We are grateful to V.L. Velichansky and V.P. Yakovlev for interest
in this work and stimulating discussions.

\end{document}